**Fair game: Urban free-ranging dogs balance resource use and risk aversion at seasonal fairs.**

Sourabh Biswas[1], Kalyan Ghosh[1], Hindolii Gope[1], & Anindita Bhadra[1*]

sb18rs107@iiserkol.ac.in, kg23rs078@iiserkol.ac.in, hindolii@yahoo.com, abhadra@iiserkol.ac.in

[1]Behaviour and Ecology Lab, Department of Biological Sciences, Indian Institute of Science Education and Research Kolkata, Mohanpur, West Bengal, India.

Abstract:

Seasonal fairs, bustling with human activity, provide a unique environment for exploring the intricate interplay between humans and free-ranging dogs. This study investigated these interactions during seasonal fairs in Nadia and Bardhaman districts, West Bengal, India. Across 13 fairground sites, we explore how human footfall and resource availability impact dog behavior, with a particular emphasis on cognitive mechanisms. Data collection occurred from December to March, comprising three sessions—initial, middle, and end—each on three randomly selected days. Employing spot census and scan sampling methods, observers documented GPS locations, sex, and behaviors of free-ranging dogs. Videos captured interactions within the fair environment, revealing cognitive processes. Our analysis revealed a notable increase in human activity during the middle phase, coinciding with a rise in dog abstractabundance. Dogs predominantly foraged, exhibited gait, and remained vigilant, their numbers positively associated with resource availability. Proximity to the fairground significantly shaped dog behavior, indicating cognitive processes in decision-making. Dogs closer to the fair demonstrated consistent behavior, likely due to immediate resource access, implying sophisticated cognitive mapping and resource utilization. Conversely, dogs from farther distances exhibited lower consistency and heightened aggression, intensifying foraging, gait, and vigilance activities, suggesting cognitive adaptations to resource scarcity and competition. These findings underscore the intricate relationship between human activity, resource availability, and the behavior and cognition of free-ranging dogs during seasonal fairs. They offer insights into the ecological dynamics of free-ranging dogs in human-influenced landscapes, emphasizing the necessity for comprehensive management strategies in urban and peri-urban environments.

**Introduction**

Anthropogenic activities greatly alter natural habitats and impact the behavior of species residing in these habitats. Over the past three decades, the rate of urbanization has been constantly increasing across the world, with the sharpest increase in cities and suburbs and concurrent decrease in rural areas being seen in Asia (Ritchie et. al, 2024).Urbanization and urban adaptation are thus subjects of scientific as well as socio-economic interest in today's world. While urbanization is a permanent phenomenon with long term implications, some anthropogenic activities cause short term perturbations in habitats, which nevertheless influence the lives of many species. Seasonal touristic activities in forests and fragile ecosystems like cold deserts, mangroves, etc.; temporary fairs, picnics, etc. in open spaces; festival gatherings and carnivals are some examples of such anthropogenic activities, which can alter the food and space availability and human flux in an area for a short period, impacting movement patterns, feeding and resting habits, biological clocks and even survival of species, especially animals.

Seasonal fairs in India represent dynamic socio-cultural events with substantial human footfall, creating transient yet profound alterations to the local environment (Jauhari & Munjal, 2015). These fairs act as major socio-cultural hubs in urban and semi-urban areas, that draw large numbers of people of all ages for a span of a few days to a relatively small area. They serve as crucibles of human interaction, culture, commerce, and festivity, engendering an intricate interplay between anthropogenic activities and the surrounding ecosystem (Cudny, 2016). A conspicuous outcome of these fairs is the sudden surge in varied food sources, catalyzed by the convergence of diverse culinary practices, street vendors, and communal feasting (Sol et al., 2013; Tribhuwan & Tribhuwan, 2003). Though fairs are a common occurrence in many parts of the world and are likely to impact and influence the eco-ethology of multiple species that live in proximity to humans in varied habitats, they have not been studied extensively from an ecological perspective. The behavior of scavenging species that can exploit human generated resources are of particular interest in such altered habitats.

Free-ranging dogs (*Canis lupus familiars*) and humans co-exist in various habitats in the Global South (Sen Majumder et al., 2014). The interactions between the two species are very complex, consisting of both positive and negative social interactions from both sides. Humans are the major source of food and shelter for these dogs, but also the biggest threat to their survival (Paul et al., 2016). These dogs show friendly and submissive gestures towards humans (Bhattacharjee et al., 2018), and beg from them for food, but instances of aggression from dogs, sometimes leading to morbidity and mortality in humans are also known (Paul et al., 2016). The problem of aggression from dogs on streets is amplified by the threat of zoonotic diseases like rabies, which is a major health problem in India and several other countries of the Global South. The dogs are key players in the scavenging community of urban ecosystems (Biswas et al., 2022), and because of their varied and constant interactions with the most dominant urban species, the humans, they are an obvious subject of interest from an ecological perspective. Moreover, understanding the nuances and the impact of anthropogenic activities on their behavior are essential for effective management strategies for urban ecosystems, to ensure co-existence and sustainability.

An earlier study on the impact of a sudden short-term surge in resources and human flux revealed that the free-ranging dogs tend to avoid the overcrowded areas, in spite of the availability of excess food. During the period of surge, the dogs tended to cluster less and remained more vigilant than at other times, suggesting that they not only avoid the sudden crowding by humans, but are also wary of them. This study was carried out by sampling each site at one time point each before, after and during an event – Durga Puja, in the state of West Bengal, India (Bhattacharjee & Bhadra, 2021). During Durga Puja, there is high flux of humans and heightened resources almost round the clock, for a period of about a week. Fairs, on the other hand, open and close at fixed times during the day over a similar span of time, and the human flux is variable during this period. Since fair grounds are pre-identified spaces, fairs provide us with the opportunity to not only explore the behavioral dynamics of free-ranging dogs during the period of the fair with respect to food and human flux surge, but also are an excellent paradigm for testing the predictions of the resource dispersion hypothesis (RDH), which was proposed to explain resource-dependent spatial distribution of scavenging species (Carr & Macdonald, 1986; Macdonald & Johnson, 2015)

In this paper, we describe a population level study carried out to understand the adaptive strategies, if any, that free-ranging dogs employ to exploit the heightened resources during seasonal fairs, while navigating the enhanced human flux. Further, we use the data from multiple such fairs to test if the spatio-temporal dynamics and behavioral patterns of free-ranging dogs in these fairs align with the predictions of RDH. The fairs were marked by a notable surge in human mobility, the presence of temporary stalls selling varied food items and different kinds of consumer goods and various entertainment options like live performances by artists, rides for children, gaming stalls, etc. Each fair attracted a substantial number of visitors who engaged in various activities and consumed large quantities of food. In this dynamic environment, dogs typically had access to food from garbage bins and through begging; however, this access might have been hindered by the increased human overcrowding. The interplay between human activities and free-ranging behavior in this setting formed the core focus of our study.

We hypothesize that the free-ranging dogs resident in the area where fairs are organized would benefit the most from the resource surge, and would maintain a low profile in terms of territorial aggression between groups to avoid negative interactions from humans. They would exploit the garbage generated at the fair grounds after the human flux reduces, towards the closing time of the fairs, thereby avoiding the peak time of human flux.

**Methodology:**

The study was conducted at 14 seasonal fairs in the Nadia, North 24 Parganas and Bardhaman districts of West Bengal, India (Fig. S1), where various annual fairs are held between the months of December to March. We selected public fairs as our primary research locations due to their regulated yet dynamic environment. An event was identified as a unique fair (with a name and identity known to the local people) that occurred at a site in a given year. Some sites could have multiple such events in a year, and typically were identified as separate fairs, with unique identities, by the local community, with a minimum of seven days gap between two such fairs. A fair could last for three days to two weeks, depending on the occasion for holding the fair.

Each event was sampled on three randomly selected days from the first to the last day of the fair. On each day, data was collected in three sessions - initial (1630 to 1800h), middle (2000 to 2130h), and the end (2230 to 0000h). During each session, an observer traversed the entire fairground along a pre-determined route, on foot, and noted the GPS location, sex, and behavior of all the free-ranging dogs sighted, using the spot census method (Sen Majumder et al., 2014b), and the behavior of each dog at the time of sighting was recorded using the instantaneous scan sampling method (Altman, 1975). Subsequently, the observer opportunistically captured multiple 5-15-minute videos of the dogs' activities at different locations within the fairground's boundary during each session, to gather further information on the behavior of the dogs, especially with respect to their interactions with humans and among themselves, during the fair. The videos were later decoded for dog behaviors;

instantaneous scans were conducted from the videos at random intervals, averaging two scans per minute. The number of scans varied based on the duration of the video in minutes. During each scan, we recorded the identity of the dogs present and their behaviors at that instant. To normalize the frequency of behaviors within a video, we divided the sum of occurrences of a specific behavior by the total number of scans conducted from that video.

The GPS coordinates of both dogs and resource points were acquired using the ©Google My Maps application on a cellular mobile phone. Following the methodology outlined in Bhattacharjee & Bhadra, (2021), human flux was recorded at a minimum of three random locations within the boundary of every fair. At each location, an observer stood and counted the number of people passing by in one minute. The average human flux was estimated for each session.

Seven days before and after the event, we conducted a comprehensive census of dogs within a 200-meter radius from the periphery of the fairground. Our primary objective was to determine the distance from which dogs were attracted to the fairground. During this census, we categorized dogs based on their behaviors, such as resting together, scouting together, or engaging in territorial fights with neighboring groups, which helped us define their group affiliations. We calculated distances from the midpoint of their territories to the periphery of the fair ground to categorise them into different fare zones. We recorded detailed descriptions of the observed dogs, including their group compositions. Additionally, we photographed the individuals for further reference in identifying the dogs at the fair.

Statistical Analysis:

All statistical analyses were conducted in R Studio (version 4.2.0) (R Core Team, 2022). To evaluate differences in human flux and abundance of dogs across event sessions, the Kruskal-Wallis rank sum test was employed. Post-hoc Dunn's tests were subsequently applied to validate specific differences among the categories. Human flux categories were classified as low, medium, and high, according to Bhattacharjee and Bhadra (2021). The behaviors observed to be performed by the dogs were categorized into foraging, gait, affiliative, aggressive, territorial, vigilance and other behaviors (Table. S1). The proportion of time spent in each behavioral category was estimated for individual dogs, normalized by their presence at a fairground on a day. We employed a Kruskal-Wallis rank sum test to analyze the variance in time spent across different behavioral categories (Table. S1) of dogs within the fairground. Subsequently, Dunn's post-hoc tests were conducted to assess pairwise differences.

The "lmer" package was employed to execute GLMM with a Poisson distribution for the number of dogs present in each event. The response variable was the count of dogs per session, and the predictor variables included resource score (Table S2), the number of entries in the fairground, and the effect of the event sessions, with unique sampling sessions as the random factor.

To compare the distance from the centroid of the fairground to the periphery and the distance between dogs and resources, we applied the Wilcoxon signed-rank test. The area around the fairground was categorized into three zones based on distance: zone a: up to 50 meters from the periphery of the fair, zone b: 50 to 150 meters from the periphery and zone c: the area beyond 150 meters from the fairground perimeter (Fig.1). Pre and post-fair surveys were conducted to check for the presence of dog groups in these zones, and to map them to the dogs recorded at the fair during the event.

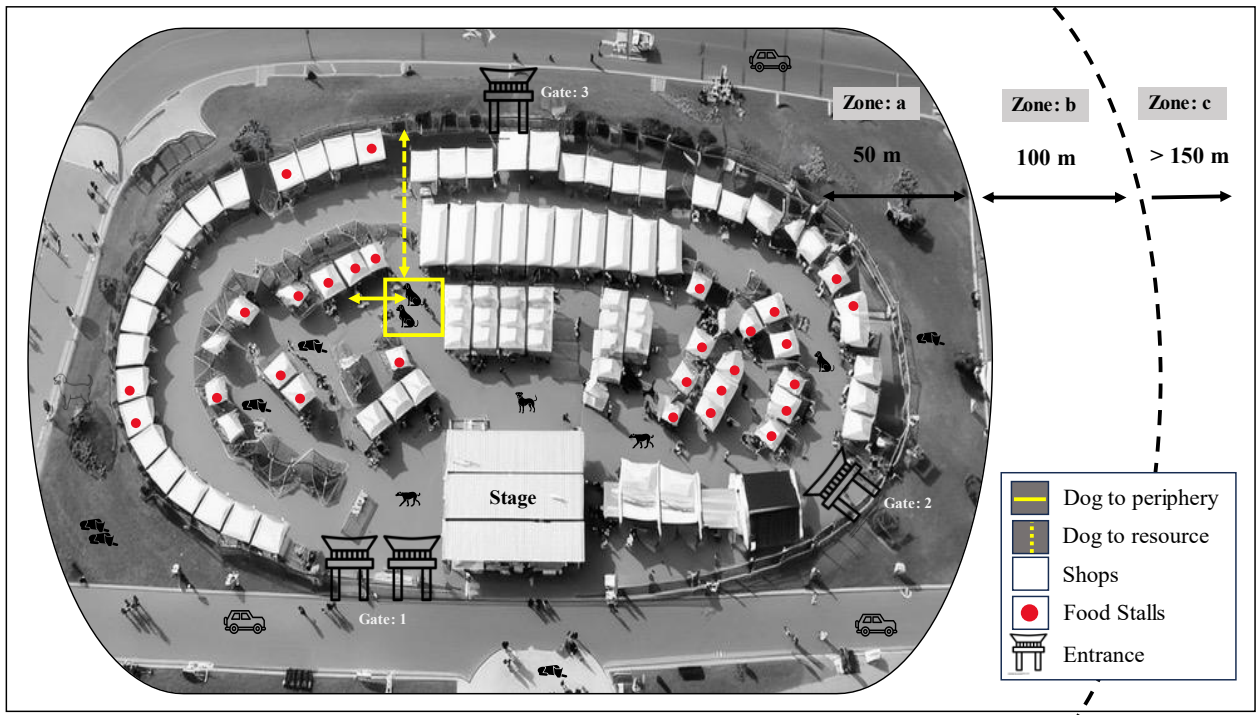

***Fig. 1: Graphical Representation of fair:*** *A 3-dimensional representation of a typical seasonal fair ground in India. There is a stage for performances, and several rows of stalls selling miscellaneous items, including food. Here, the food stalls are demarcated by red dots. This diagram shows a fair with three gates. Representative visitors and dogs are marked within and around the fair. There is a road marked on one side. The area around the fairground was categorized into three zones based on distance: zone a: up to 50 meters from the periphery of the fair, zone b: 50 to 150 meters from the periphery, and zone c: the area beyond 150 meters from the fairground perimeter. In yellow, the method used for calculating the distance from a focal dog to the nearest resource and the nearest point on the periphery are shown.*

The consistency of dogs in a specific location is defined by how frequently they are present during sampling periods at the study site. This is calculated by dividing the number of times a dog is sighted at the fair by the total number of sampling bouts. Dogs were then categorized into three levels of consistency: high (> 0.66), moderate (0.33-0.66), and low (< 0.33), based on their observed presence at the fair. Pairwise chi-square tests (a vs. b, b vs. c, and a vs. c) were performed to compare the distribution of dogs from the three zones sighted at the fair.

In our study, we employed two statistical models in 'R' to assess the consistency of dogs: the Generalized Linear Mixed-Effects Models (GLMM) and the Cumulative Link Mixed Models (CLMM). The GLMM, with a beta distribution, was implemented using the 'glmmTMB' package (Bolker, 2019). This model was used to evaluate the overall consistency of dogs. In contrast, the CLMM was applied to assess both session-wise and daily consistency. These models were chosen due to the discrete nature of our response variable (Table 4) and was implemented using the 'ordinal' package (Haubo et al., 2022). For all the models, we considered 'gender' and 'zone' as predictors and included 'place' as a random factor.

We also conducted a separate analysis for the Kalyani area, which hosted three consecutive fairs at the same fairground over approximately two months. In this context, we used the GLMM with beta distribution to assess session-wise consistency and the CLMM to evaluate overall and daily consistency. In the CLMM, the response variables were ordered due to their discrete nature (see ESM). For all models, 'gender' and 'zone' were considered as predictors. Additionally, 'event' was included as a predictor variable for the session model.

Due to the small sample size in different fair category zones, zones a and b were combined into 'zone 1', representing the vicinity (within 150 meters), while zone c was categorized as 'zone 2', representing areas away from the fairground (more than 150 meters). Kruskal-Wallis rank sum tests were used to compare the level of consistency across different behavior categories. We categorized dogs into "consistency-zones" based on their consistency levels (high, moderate, and low) and their respective zones (a, b, c). For instance, a dog with high consistency from zone 'a' would be labelled as 'high_a,'. This process resulted in a total of nine consistency-zone categories. The time spent in different behaviors by dogs belonging to these consistency-zone classes were compared using Kruskal-Wallis rank sum tests.

**Results:**

**Temporal and spatial dynamics of human and dog presence at the fair:**

The dynamics of human and dog presence at the fair showed variations throughout the day. Human activity fluctuated significantly over time (Kruskal-Wallis rank sum test; $\chi^2 = 59.29$, df = 2, p < 0.001), with Dunn's test highlighting distinct differences between stages (p < 0.05). Specifically, human flux peaked during the middle phase (85.09 ± 38.84), contrasting with lower levels observed during the initial (29.09 ± 18.73) and final (19.84 ± 7.89) periods (Fig. S3a). The number of dogs present showed variation across the fair's duration (Kruskal-Wallis test: $\chi^2 = 6.2706$, df = 2, p < 0.05), with Dunn's test indicating differences between phases. Dog numbers peaked during the final phase (7.63 ± 5.06), surpassing counts during both the initial (5.44 ± 3.99) and middle (5.02 ± 3.68) periods (Fig. S3b).

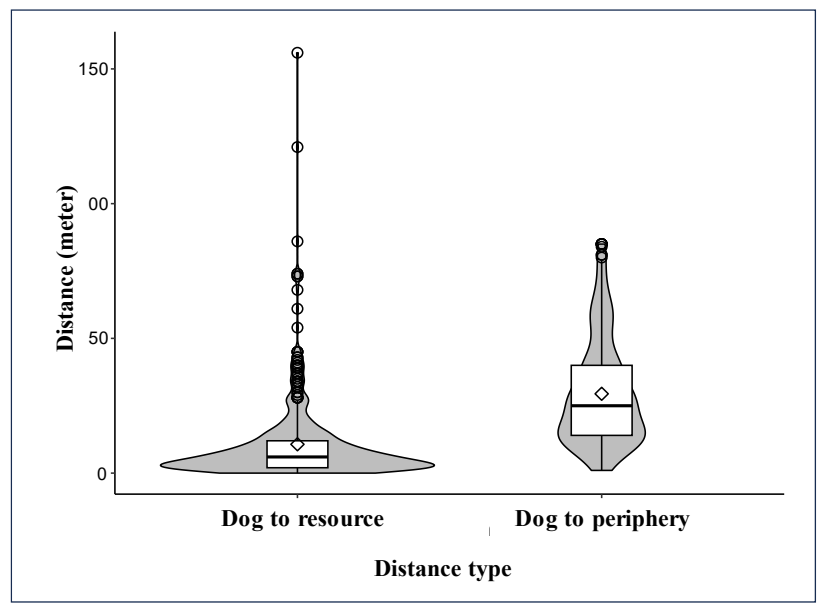

*Fig. 2: Spatial location of dogs in the fair: A box plot embedded within a violin plot describes the distance of dogs from the nearest resource and the nearest point at the periphery of the fair.*

The distance between dogs and resources (10.69m ± 15.08m) were significantly lower than the distance between dogs and the boundary of the fair (29.43m ± 19.41m) (Wilcoxon signed-rank test; V = 11972, p < 0.001; Fig.2). This suggests that the dogs preferentially stayed close to the resource sites within the fairs.

Generalized linear mixed-effects model (GLMM) analysis unveiled a significant positive impact of resources on dog abundance (Z = 3.461, p < 0.001). Additionally, the presence of three or four entry points into the fair exhibited positive associations with dog abundance (Z = 2.441 and 2.774 respectively; p < 0.05), as opposed to one or two entry points (Table. S3).

**Free-ranging dog behavior at the fairs:**

The behaviors observed to be performed by the dogs were categorized into foraging, gait, affiliative, aggressive, territorial, vigilance and other behaviors (Banerjee & Bhadra, 2022)). The proportion of time spent in each behavioral category was estimated for individual dogs, normalized by their presence at a fairground on a day. This provided an estimate of the time-activity budgets of the dogs within the fair. The behaviors of dogs showed significant variation across different categories (Kruskal-Wallis rank sum test; $\chi^2$ = 279.46, df = 6, p < 0.001). Subsequent Dunn's post-hoc tests for pairwise comparisons (p < 0.05) revealed that dogs spent the majority of their time engaging in foraging (0.44 ± 0.22), followed by gait (0.14 ± 0.09), and vigilance (0.14 ± 0.08), compared to other behavior categories (Fig. S4).

**Dog distribution and consistency at the fair:**

The area outside the fair boundary was demarcated into three zones (Fig. 1) and the representation of dogs from the three zones within the fair varied significantly ($\chi^2$ = 30.515, df = 2, p < 0.001). Pairwise chi-square tests revealed that the number of dogs from zones 'a' and 'b' were similar ($\chi^2 = 0.82051$, df = 1, p = 0.365), while both surpassed the count from zone 'c' ($\chi^2 = 16.03$ for 'b' vs 'c' and $\chi^2 = 23.516$ for 'a' vs 'c'; df = 1, p < 0.001 for both).

The analysis revealed a significantly lower consistency of dogs coming from zone 'c' across fair sessions in event session (CLMM, Z = -2.402, p < 0.05), in a day at the fair (CLMM, Z = -4.857, p < 0.001) and overall (GLMM, Z = -3.892, p < 0.001) across the fairs. Similarly, gender did not influence dog consistency across sessions, and days (p > 0.05) but in the overall categories male dogs exhibited a significantly lower consistency (p > 0.05) compared to females (Tables S4-S6).

In the Kalyani fairs, the analysis revealed gender-related influences on dog consistency across various parameters. Male dogs exhibited a significantly lower consistency during event sessions (GLMM, Z = -3.148, p < 0.001), on a daily basis (CLMM, Z = -2.794, p < 0.01) compared to females across fairs. However, the difference in overall consistency between male and female dogs was not significant (p > 0.05). Similarly, dogs from zone '2' also showed a lower consistency during event sessions (GLMM, Z = -3.056, p < 0.01), on a daily basis (CLMM, Z = -3.919, p < 0.001) and overall (CLMM, Z = -1.986, p < 0.05) compared to those from zone '1'. (Tables S7-S9).

**Effects of consistency levels and zones on dog behavior:**

Kruskal-Wallis tests revealed significant effects of both consistency levels and zones on scavenging, aggressive , begging, vigilance, and gait behaviors.

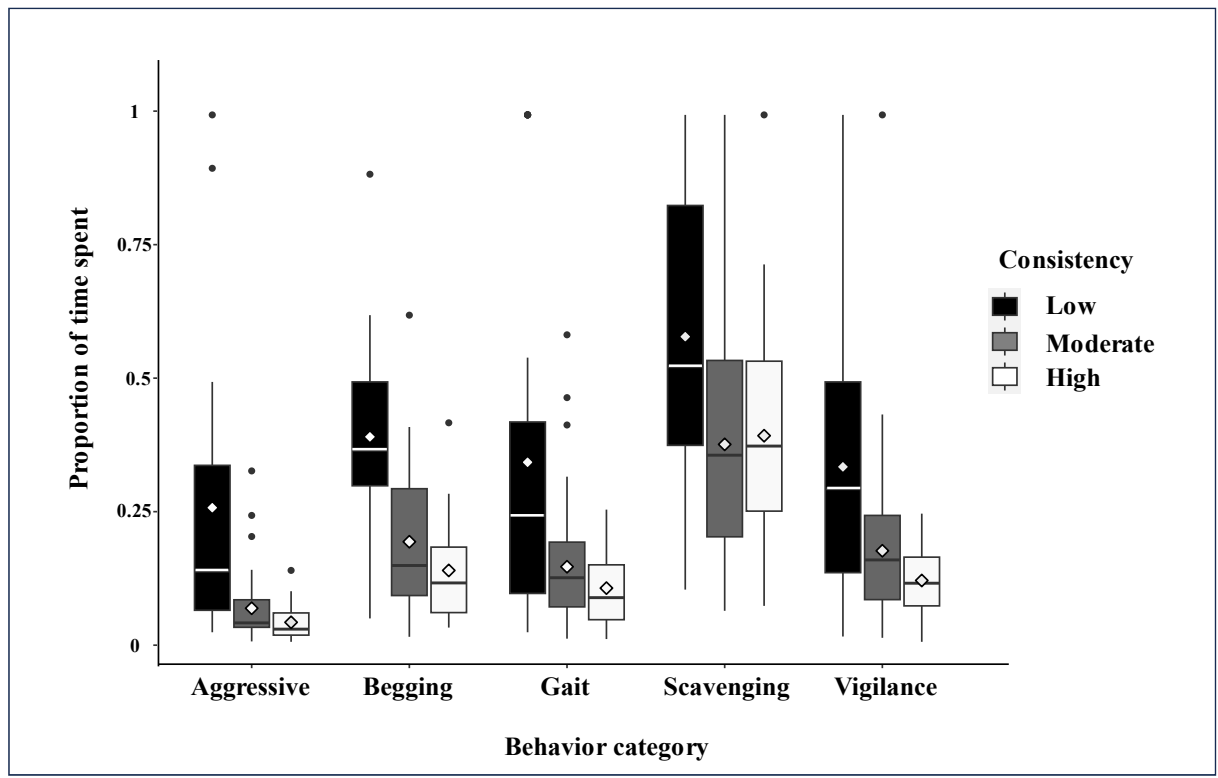

*Fig. 3:* A box and whiskers plot describing the comparison of different behaviors recorded at fairs by dogs of different consistency levels. The line within each box represents the median value, the open diamond marks the mean, the rectangle shows the 25th and 75th quartile of the data and the whiskers represent the data range.

For scavenging behavior, significant overall effects were found for both consistency levels ($\chi^2$ = 15.524, df = 2, p < 0.001) (Fig. 3) and consistency-zones ($\chi^2$ = 20.661, df = 8, p < 0.01) (Fig. S5). Dogs categorized as 'low' consistency exhibited significantly more scavenging than those classified into 'mod' and 'high' levels, while dogs in the 'low_a' and 'low_c' categories scavenged significantly more than those in 'high_b', 'high_c', and 'mod_b' zones.

Significant effects were observed for both consistency levels ($\chi^2$ = 22.38, df = 2, p < 0.001) (Fig. 3) and consistency-zones ($\chi^2$ = 29.645, df = 8, p < 0.001) (Fig.S6) for aggressive behavior. Dogs in 'low' consistency-zones (low_a & low_c) displayed higher aggression compared to 'high_b' and 'mod_a' category dogs, with dogs in the 'mod_a' category showing lower aggression than those in 'low_a' and 'low_c' categories. Similarly, dogs in the 'low' consistency level exhibited more aggression than those in 'mod' and 'high' levels.

Likewise, for begging behavior, significant effects were found for both consistency levels ($\chi^2 = 24.402$, df = 2, p < 0.001) (Fig. 3) and consistency-zones ($\chi^2 = 29.147$, df = 8, p < 0.001) (Fig. S7). Dogs in 'low' consistency-zone categories ('low_a', 'low_b', 'low_c') exhibited higher begging compared to 'high_b', 'high_c', and 'mod_a' categories. Additionally, dogs in the 'mod_a' category showed higher begging compared to 'high_c' and 'mod_b' categories. Dogs with 'low' consistency level showed more begging compared to 'mod' and 'high' levels, while dogs in the 'mod' level exhibited more begging compared to the 'high' level.

Furthermore, significant effects were observed for both consistency levels ($\chi^2 = 17.971$, df = 2, p < 0.001) (Fig. 3) and consistency-zones ($\chi^2 = 21.252$, df = 8, p < 0.01) (Fig. S8) in vigilance behavior. Dogs in the 'low_b' category exhibited higher vigilance compared to 'high_b' category, with a similar trend observed for 'low_c' compared to 'high_c' and 'mod_c' categories. Regarding consistency levels, dogs at the 'low' level displayed more vigilance compared to 'mod' and 'high' levels.

Similarly, for gait behavior, significant effects were found for both consistency levels ($\chi^2 = 18.466$, df = 2, p < 0.001) (Fig. 3) and consistency-zones ($\chi^2 = 23.842$, df = 8, p < 0.001) (Fig. S9). Dogs in the 'low_b' category exhibited higher gait behavior compared to 'high_b' and 'mod_b' categories, with a similar trend observed for 'low_c' compared to 'high_c' and 'mod_c' categories. Regarding consistency levels, dogs in the 'low' level displayed more gait behavior compared to 'mod' and 'high' levels.

**Discussion:**

Our investigation into the impact of human footfall and sudden resource generation on dog distribution and behavior at seasonal fairs revealed distinct patterns. Human activity peaked during the middle of the fair's operating hours, while the dog numbers peaked when human abundance reduced at the fairs after a peak. The peak in human activity during the middle period of the fair may be influenced by various factors such as the timing of popular events, attractions, or performances occurring during this timeframe. As the fair draws to a close towards late evening, there is typically a reduction in human activity and a subsequent decrease in the intensity of human presence within the fairgrounds. As visitors leave and vendors begin to pack up, food scraps and leftovers may become more accessible to dogs, drawing them in larger numbers to the fairgrounds. Moreover, as the fair winds down, there are typically fewer disturbances or deterrents for dogs, such as loud noise, unfriendly people, etc. The reduction of potential stressors and more available space within the grounds may further encourage dogs to roam more freely and congregate in larger numbers towards the final hours of the fair. These observations align with Bhattacharjee and Bhadra (2021), where a negative impact of a surge in human flux was observed on dog abundance during Durga Puja, an annual socio-religious event in West Bengal.

The proximity of the original territories of dogs played an important role in determining their presence and behavior at the fairgrounds. Dogs from original territories close to the fairgrounds exhibited a high level of consistency of being present at the fair, likely due to their immediate access to resources. In contrast, dogs from farther areas displayed lower consistency and heightened aggression, engaging more intensely in foraging, gait, and vigilance activities while present at the fairs, as compared to the more consistent dogs. Thus, it appears that while the dogs that are resident of areas in the vicinity of the fairs explore the heightened resources at a slow and steady pace, defending them actively, those coming in from farther distances use a "grab and go" strategy to maximise their relatively shorter time spent at the fairs.

The behavioral profiles of the dogs at the fair provided interesting insights. Firstly, the heightened foraging behavior indicates that dogs are actively searching for and exploiting available resources within their environment. This confirms that it is the available resources that have attracted them to the fairs, for some, in spite of the risks associated with crossing territories of other dogs to reach the site. This aligns with the Resource Dispersion Hypothesis (RDH), as it suggests that dogs are responding to the spatial distribution of resources by focusing their efforts on areas where resources are more abundant (Carr & Macdonald, 1986). Dogs have been found to play a key role in the scavenging network of anthropogenic food sources and monopolise the resource over other scavenger species (Biswas et. al., 2024). In the fairgrounds we focused solely on the dogs, and found them to be efficiently exploiting these transient yet abundant resources.

Similarly, the increased gait behavior, which includes all forms of locomotion like walking, running, ambling, etc., reflects active exploration of the area to locate resources. Dogs may adjust their gait depending on their level of alertness and interest in their surroundings, facilitating their navigation through the fairgrounds in search of food, water, or other resources. Additionally, heightened vigilance behavior indicates that dogs are monitoring their surroundings for potential threats or opportunities. This behavior is crucial for navigating crowded and dynamic settings effectively, ensuring the safety of the dogs and facilitating their interactions with the surrounding stimuli, including potential resources. The increased vigilance observed in our study re-affirms the observations by Bhattacharjee and Bhadra (2021) during Durga Puja. Overall, the observed behaviors suggest that dogs are actively responding to the spatial distribution of resources within their environment, consistent with the predictions of RDH. Additionally, the association between the number of entry points into the fair and dog abundance further supports this trend, suggesting that easier access to resource-rich areas facilitates increased dog presence.

The higher consistency observed in female dogs compared to males at the Kalyani fairs may stem from various factors. Female dogs often exhibit more stable social behaviors and are less prone to territorial disputes, which could contribute to their ability to maintain consistent patterns within the fairground (Borchelt, 1983; Pérez-Guisado et al., 2006). Additionally, females may be more responsive to environmental cues and less easily distracted, leading to greater consistency in their behavior over time (D'Aniello et al., 2021). Thirdly, the dog population in the city of Kalyani might be female-biased, leading to more female dogs gaining access to the fairground. These differences in behavior between genders highlight the importance of considering gender dynamics in understanding the behavior of free-ranging dog populations in urban environments.

Our spatial distribution analysis revealed a distinct pattern in dog distribution; the dogs were observed to be closer on average to the resource-rich areas within the fairs, as compared to the periphery of the fairgrounds. This spatial pattern aligns with the predictions of the RDH (Macdonald & Johnson, 2015; Mbizah et al., 2019; Valeix et al., 2012), suggesting that dogs strategically position themselves closer to areas with abundant resources to maximize their access.

(Calcagno et al., 2006) proposed that predictable resources are more easily exploited by species with large foraging radii, while unpredictable resources are exploited in larger numbers by local dominant species, reflecting a competition-colonization trade-off hypothesis that explains species coexistence. This concept elucidates the response observed in the significant number of dogs arriving from distant territories and sharing the same space to exploit the resources at the fairs. Since fairs predominantly occur at night, it may restrict the opportunity for other scavenger species within the anthropogenic food subsidies guild (Biswas et al., 2022) to access the resources, allowing dogs, as dominant species, to monopolize them. However, these other species may scavenge the remaining resources during daylight hours, demonstrating the temporal dynamics of resource exploitation within urban environments.

While our study didn't directly address group formation (Johnson et al., 2002), the increased dog abundance near resources implies a potential temporary aggregation due to shared interests, including dogs from different territories such as those near the fairground and those from farther away. Such aggregation is not expected or observed usually for free-ranging dogs in the urban environment. Social interactions like affiliation and play were found to significantly influence the observed patterns, suggesting that these aggregations were not permanent social groups. Thus, the dogs from different territories were co-existing and co-feeding within the fairs, occasionally engaging in aggressive interactions, over a short temporal scale. This emphasizes the role of resource availability in shaping the spatial ecology and behavior of free-ranging dog populations in urban environments, as proposed by RDH.

By actively exploring their new surroundings, animals gather information about available food sources, shelter, and potential competitors or predators (Krebs, 1982; Russell et al., 2010; Smith et al., 2009). This exploratory behavior allows them to make informed decisions about where to focus their efforts for resource acquisition. Foraging animals can choose between cautious or riskier tactics depending on the perceived threat and the benefits of foraging, leading to behavioral decisions that balance risk and reward (Gillette et al., 2000). Animals may alter their foraging behavior to maximize food intake while minimizing exposure to perceived risks, showcasing adaptive responses to environmental challenges (Dwinnell et al., 2019). In the fairs, dogs coming from far away might be employing efficient foraging strategies in unfamiliar environments by targeting high-value food sources and optimizing their efforts to maximize food intake while minimizing energy expenditure. The dogs from the distant territories exhibited more vigilance, which is crucial for detecting potential threats and opportunities. This heightened vigilance, often observed in animals in unfamiliar environments due to predation risk, allows them to quickly identify and respond to changes in their surroundings, such as the presence of other animals, human activity, or the availability of food resources (Bøving & Post, 1997; Hume et al., 2019; Pascual et al., 2014). Such adaptive behaviors enhance their ability to thrive in dynamic and unpredictable conditions within the fairground.

Aggressive behaviors, exhibited by dogs from distant locations, could be a competitive strategy employed to secure access to limited resources within a fairground (Jacobs et al., 2018). In environments where resources are scarce or highly contested, heightened aggression may provide individuals with a competitive advantage, ensuring preferential access to essential resources and potentially enhancing their chances of survival and reproductive success (Farhoody et al., 2018). Studies have shown that aggression can play a crucial role in competitive interactions among animals, with more pronounced aggression observed in populations facing higher competitive pressures (Brown, 1964; Wood et al., 2014). So, the heightened aggression displayed by dogs from distant locations may indeed serve as a competitive strategy to secure limited resources within a fairground.

Animals, including dogs, may exhibit more effective foraging strategies in unfamiliar environments by concentrating their efforts on areas with the highest probability of resource availability (Kie, 1999). This optimization strategy involves targeting high-value food sources, such as discarded food scraps or leftovers, while minimizing energy expenditure to maximize food intake (Bøving & Post, 1997; Kie, 1999). Animals may adapt to novel environments by employing flexible foraging strategies that utilize various environmental cues (Gilmour et al., 2018). Additionally, animals might modify their foraging behavior to efficiently exploit prey that dynamically changes in density and availability over time (Akiyama et al., 2019). Foraging behavior is influenced by natural selection, and animals adjust their strategies based on the predictability of the environment (Kilpatrick et al., 2021).

From an urban management perspective, the findings stress the need for tailored strategies that account for spatial variations in resource availability and human-dog interactions. By implementing targeted interventions, policymakers can effectively manage conflicts and foster coexistence between humans and free-ranging dogs in urban areas. Moreover, the study underscores the pivotal role of resource management in shaping animal behavior and spatial distribution within urban settings, advocating for sustainable management practices that support both human and animal welfare.

In the realm of urban adaptation, the research illuminates how animals, such as dogs, adapt to dynamic environments like seasonal fairs in close proximity to humans. By examining dogs' responses to human activities and available resources, the study provides insights into the adaptive strategies employed by urban wildlife populations. Moving forward, future research should delve deeper into the mechanisms driving human-animal interactions in urban environments and assess the efficacy of management interventions. By integrating ecological principles with urban planning and management strategies, a more sustainable and harmonious urban environment can be created for both humans and wildlife.

**Acknowledgements:**

The authors would like to acknowledge the Indian Institute of Science Education and Research Kolkata for providing infrastructural support. SB would like to thank the University Grants Commission, India for providing him doctoral fellowship. We also like to thank Janki Amal Grant for supporting the field work.

**Author contributions**

SB, KG and HG carried out the field work. SB and KG curated the data, SB carried out the data analysis. AB conceived the idea, supervised the work, reviewed and edited the manuscript.

**Data availability statement**

Data supporting the results will be archived.

**Conflict of Interest Information**

Authors don't have any conflict of interest.

**Supplementary materials:**

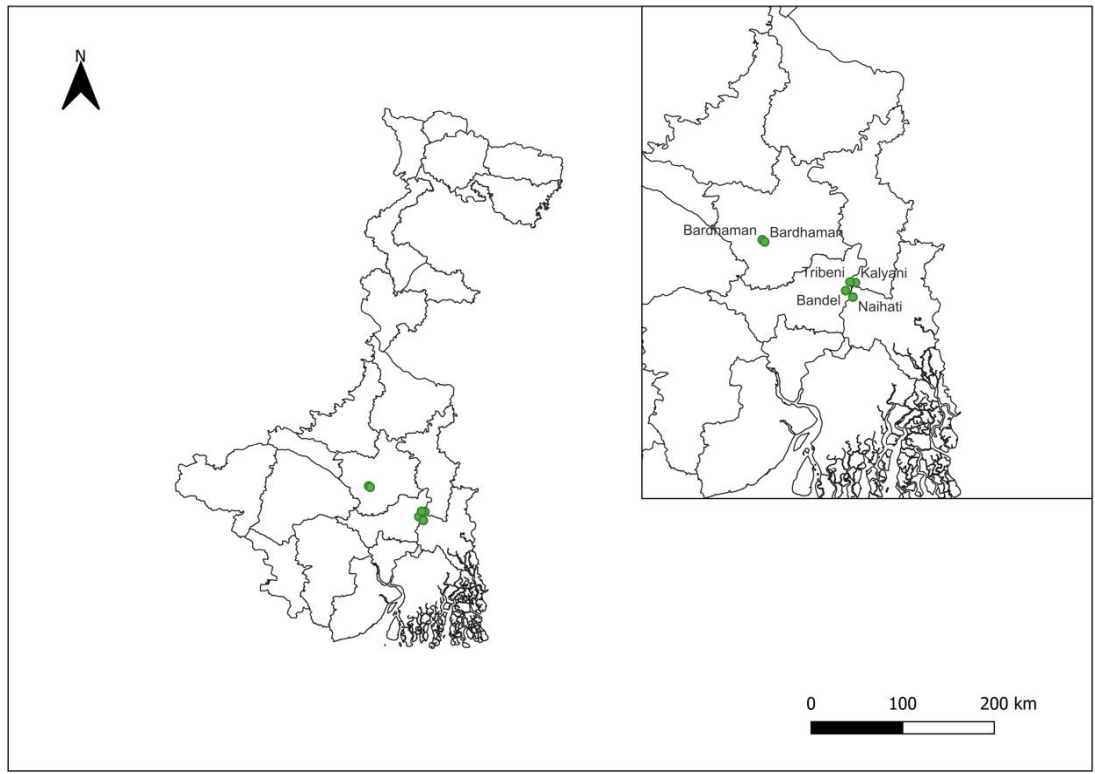

**Fig. S1: Study sites:** A map of West Bengal, India, showing the locations of 14 fairs. We covered 7 fairs in Kalyani and one fair each in Bandel, Tribeni, and Haringhata within the Nadia district, as well as one fair in Naihati in the North 24 Parganas district. Additionally, we covered 3 fairs in Bardhaman town in the Bardhaman district.

**Table S1: Behavior categories**: The behaviors observed to be performed by the dogs were categorized as per Banerjee & Bhadra, (2022).

| Behavior Category | Behaviors |
|---|---|
| Affiliative | Tail wag, Being offered food, Showing Affiliation |
| Aggressive | Being Scolded, Receive threat, Showing aggression, Barking, Receive kick, Growling, Snarling, Howling, Chase other, Chase dog, Being chased, Refuse food, Loose food, Chase human, Being barked at, Dog gets beaten. |
| Begging | Begging and or soliciting food |
| Gait | Trot, Walk, Crawling, Run, Ambling, Walk around |
| Maintenance | Groom, Scratching, Yawning, Lick self, Urinate, defecate, panting, walk in circle, shivering, Roll, Eating grass, Removing insect |
| Mating Affiliative | Investigate urine spot, Genital Sniffing, investigate another dog, Showing play bow |
| Mating Aggression | Throw the male off-balance, Aggressive toward approaching partner, Crouch down or moving away during mount attempt |
| Mating Related Other | Sniffing urine, licking urine, Nape bite, mount, try to clasp, being tried to clasp, Running together |
| Others | Follow dog, Scratching ground, Being approached, Follow human, Arching back, Dogs are called by human, Shaking head, Approach a stationary human, Turning head, Stand up, Nudge, Standing, Masturbate, Approach dog, Being followed |
| Play Related | Mock bite, Receive mock bite, Nibble dog, Being nibbled, Showing play bow, play, object play |
| Resting | Sleep, Siting, gazing, laid down, Lazing, Sit up, Curl up, move while sleeping |
| Scavenging | food search, Sniffing around, eating, drink water, Licking quid, inspect object, carry food, Sneaking, Chewing |
| Submissive | Tail droop, Show submissive behavior |
| Territorial | Marking, Overmarking |
| Vigilance | watching, watching human, watching dog, Watching her pups, Alert |

**Table S2: Resource description**: Resource score was given in the scale of 1-5 based on the quality and expected calorigenic value of the food item.

| Sl No. | Description | Score |
|---|---|---|
| 1. | Bhelpuri, Papri- Chat, Ghugni, Phucka, Mathura Cake, Potata, Momo, Ice- cream, Popcorn, Jilapi or Sweet shop, Pithe Shop, Tea shop | 1 |
| 2. | Small roll centre, Chop Shop (both veg & non veg type) | 3 |
| 3. | Big restaurant including fast food and Chicken and fish item shop | 5 |

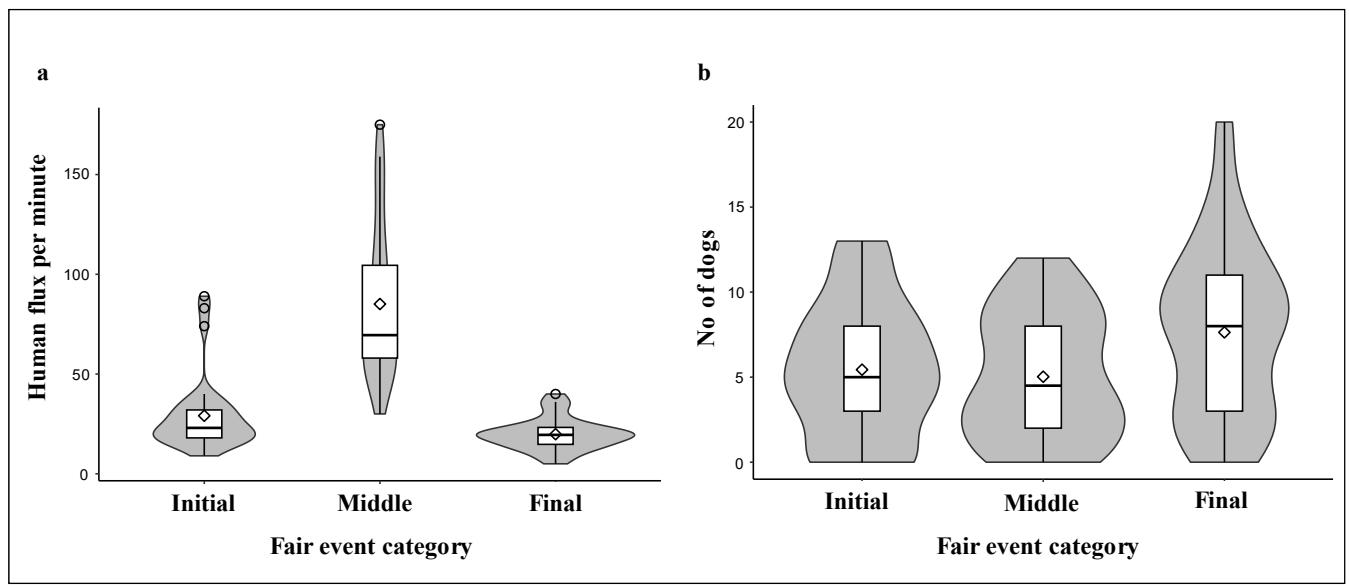

**Fig. S3:** Box plots embedded within violin plots explaining **a: Human flux**: the human flux at different times of the fair; **b: Abundance of dogs:** the abundance of free-ranging dogs at different times of the fair.

**Table. S3:** GLMM analysis describing the impact of resources, entry point and event time on dog abundance in fair.

**Formula:** Abundance of dogs in a session ~ resource score + No of entry point in a fair + Event time of the fair (initial/middle/final).

Random Factor: Location of the fair

| Fixed Effects | Estimate | Std. Error | z value | Pr(>|z|) |
|---|---|---|---|---|
| (Intercept) | 1.53842 | 0.12175 | 12.636 | < 2e-16 |
| Resource | 0.19689 | 0.05689 | 3.461 | 0.000539 |
| No_of_entry_3 | 0.37329 | 0.15292 | 2.441 | 0.014642 |
| No_of_entry_4 | 0.36564 | 0.13179 | 2.774 | 0.00553 |
| Event_initial | -0.22263 | 0.10668 | -2.087 | 0.036904 |
| Event_middle | -0.17275 | 0.10147 | -1.702 | 0.088665 |

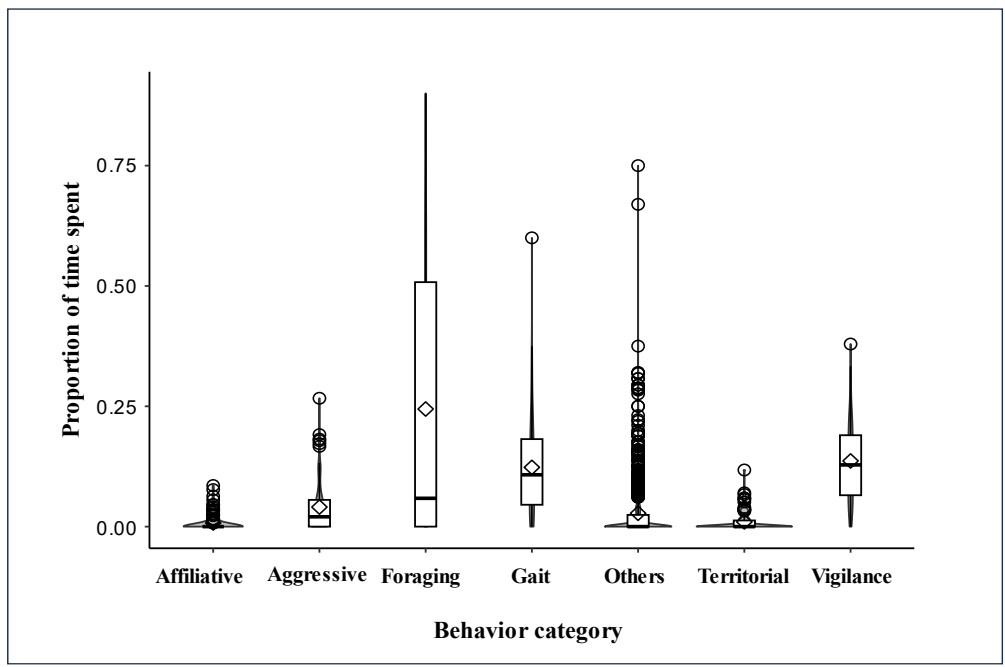

**Fig. S4:** A box plot embedded within a violin plot describing the comparison among behaviors observed to be performed by the dogs at the fairs.

**Table S4: Consistency over session:**

The Cumulative Link Mixed Models (CLMM) analysis was performed to assess the impact of gender, zones of the fair on dog consistency over sessions at the fair. The consistency of dogs was ordered due to its discrete distribution, with values assigned as follows: 0.33 as 1, 0.5 as 2, 0.67 as 3, and 1 as 4, resulting in an ordered scale from 1 to 4.

**Formula:** Ordered consistency of dogs in a session (1/2/3/4) ~ gender (male/female) + zones of the fair (a/b/c).

Random Factor: Location of the fair

| Coefficient | Estimate | Std. Error | z value | Pr(>|z|) |
|---|---|---|---|---|
| Gender_male | -0.0985 | 0.2281 | -0.4318 | 0.66585610 |
| Zone_b | 0.1541 | 0.3142 | 0.4906 | 0.62372233 |
| Zone_c | -0.7278 | 0.3030 | -2.4021 | 0.01629969 |

**Table S5: Consistency over days:**

The Cumulative Link Mixed Models (CLMM) analysis was performed to assess the impact of gender, zones of the fair on dog consistency over days at the fair. The consistency of dogs was ordered due to its discrete distribution, with values assigned as follows: 0.33 as 1, 0.5 as 2, 0.67 as 3, and 1 as 4, resulting in an ordered scale from 1 to 4.

Formula: Ordered consistency of dogs in days (1/2/3/4) ~ gender (male/female) + zones of the fair (a/b/c).

Random Factor: Location of the fair

| Coefficient | Estimate | Std. Error | z value | Pr(>|z|) |
|---|---|---|---|---|
| Gender_male | -0.3198 | 0.2276 | -1.4050 | 0.160023 |
| Zone_b | -0.7352 | 0.3131 | -2.3481 | 0.018871** |
| Zone_c | -1.5877 | 0.3269 | -4.8570 | 1.1921e-06*** |

**Table S6: Consistency overall:**

GLMM analysis was performed to assess the impact of gender, zones of the fair on dog overall consistency at the fair.

**Formula:** Consistency of dogs in a fair ~ gender (male/female) + zones of the fair (a/b/c).

Random Factor: Location of the fair

| Coefficient | Estimate | Std. Error | z value | Pr(>|z|) |
|---|---|---|---|---|
| (Intercept) | -0.43803 | 0.26608 | -1.646 | 0.0997 |
| Gender_male | -0.42287 | 0.18131 | -2.332 | 0.0197** |
| Zone_b | 0.08939 | 0.27131 | 0.329 | 0.7418 |
| Zone_c | -0.96511 | 0.24797 | -3.892 | 9.94e-05*** |

**Table S7: Session consistency of Kalyani**

GLMM analysis describe the impact of gender, zones of the fair and event time on dog consistency over session in Kalyani fairs.

**Formula:** Consistency of dogs in a session ~ gender (male/female) + zones of the fair (zone 1/ zone 2) + Event time of the fair (initial/middle/final).

Random Factor: Location of the fair

| Coefficient | Estimate | Std. Error | z value | Pr(>|z|) |
|---|---|---|---|---|
| (Intercept) | -0.19545 | 0.20287 | -0.963 | 0.33535 |
| Gender_male | -0.50738 | 0.16116 | -3.148 | 0.00164** |
| Zone_2 | -0.52191 | 0.17078 | -3.056 | 0.00224 ** |
| Event_initial | -0.09216 | 0.18765 | -0.491 | 0.62332 |
| Event_middle | 0.00695 | 0.19021 | 0.036 | 0.97085 |

**Table S8: Consistency of Days at Kalyani:**

The Cumulative Link Mixed Models (CLMM) analysis was performed to assess the impact of gender, zones of the fair, and event time on dog consistency over days at the Kalyani fairs. The consistency of dogs was ordered due to its discrete distribution, with values assigned as follows: 0.33 as 1, 0.5 as 2, 0.67 as 3, and 1 as 4, resulting in an ordered scale from 1 to 4.

**Formula:** Ordered consistency of dogs in days (1/2/3/4) ~ gender (male/female) + zones of the fair (zone 1/ zone 2).

Random Factor: Location of the fair

| Coefficient | Estimate | Std. Error | z value | Pr(>|z|) |
|---|---|---|---|---|
| Gender_male | -1.0130 | 0.3625 | -2.7942 | 0.00520268** |
| Zone_2 | -1.5124 | 0.3859 | -3.9196 | 8.8713e-05*** |

**Table S9: Consistency Overall Kalyani:**

The Cumulative Link Mixed Models (CLMM) analysis was performed to assess the impact of gender, zones of the fair, and event time on dog consistency over days at the Kalyani fairs.

The consistency of dogs was ordered due to its discrete distribution, with values assigned as follows: (0.04 & 0.09) as 1; (0.14 & 0.19) as 2; (0.23 & 0.28) as 3; (0.33 & 0.42) as 4 and (0.57 & 0.61) as 5, resulting in an ordered scale from 1 to 5.

**Formula:** consistency of dogs in a fair ~ gender (male/female) + zones of the fair (zone 1/ zone 2).

Random Factor: Location of the fair

| Coefficient | Estimate | Std. Error | z value | Pr(>|z|) |
|---|---|---|---|---|
| Gender_male | -1.1424 | 0.6224 | -1.8354 | 0.066443 |
| Zone_2 | -1.3147 | 0.6619 | -1.9862 | 0.047010* |

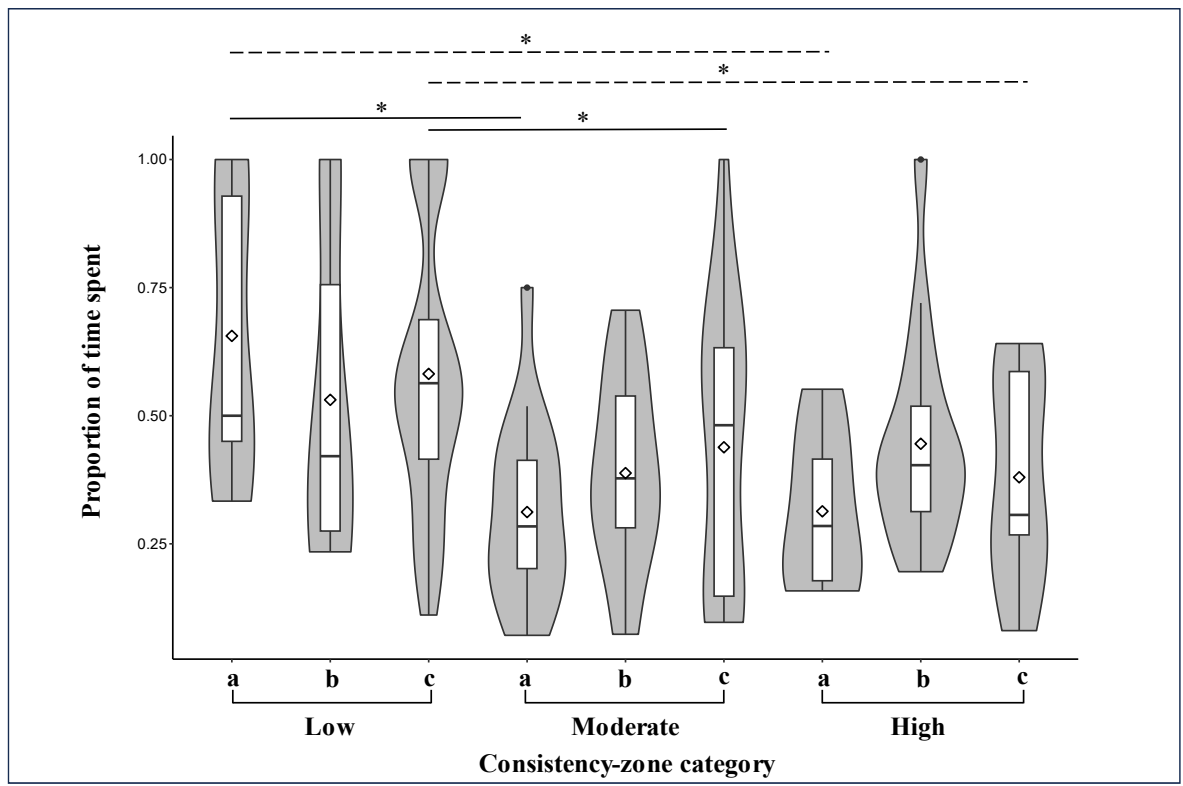

**Fig S5: Scavenging behavior:** The embedded box plot within the violin plot compares dogs' scavenging behavior across different consistency zones. The black line represents the median

value, the open diamond indicates the mean, the rectangle shows the interquartile range (25th to 75th percentiles), and the whiskers represent the data range. The violin shape illustrates the density of the data distribution.

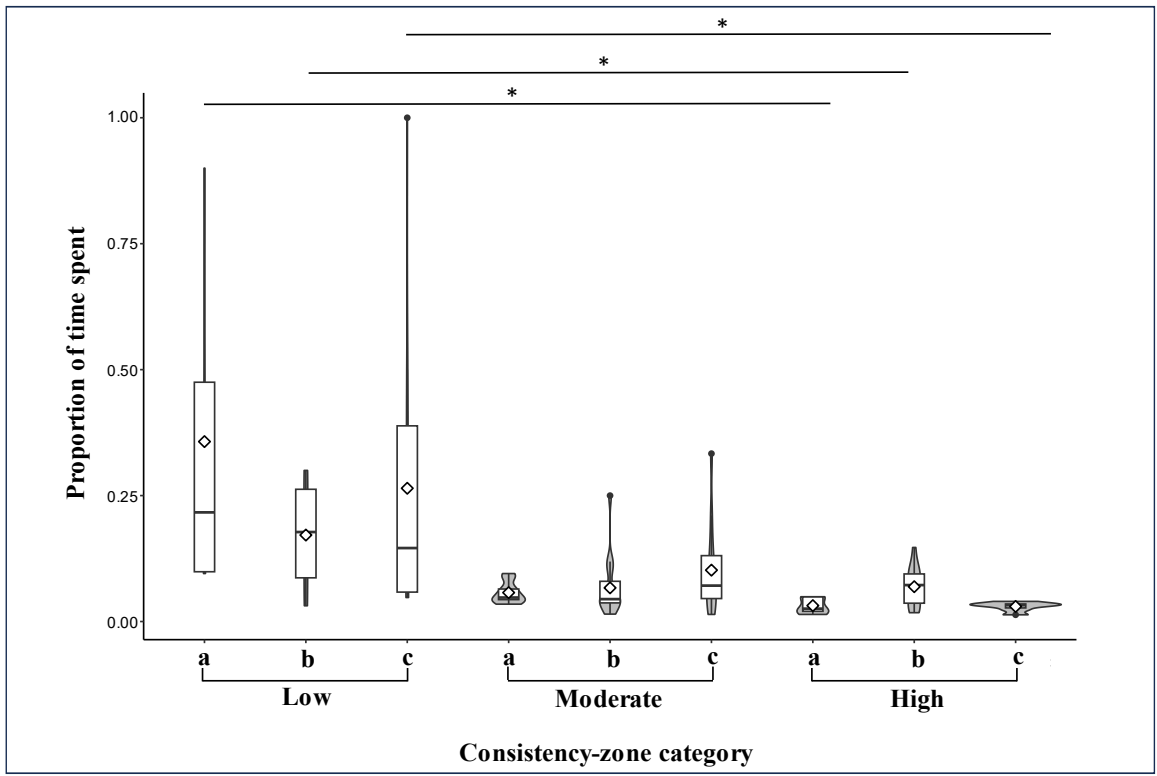

**Fig S6: Aggressive behavior:** Box plot embedded within violin plot describing the comparison among dogs from different consistency zones for aggressive behvaiour.

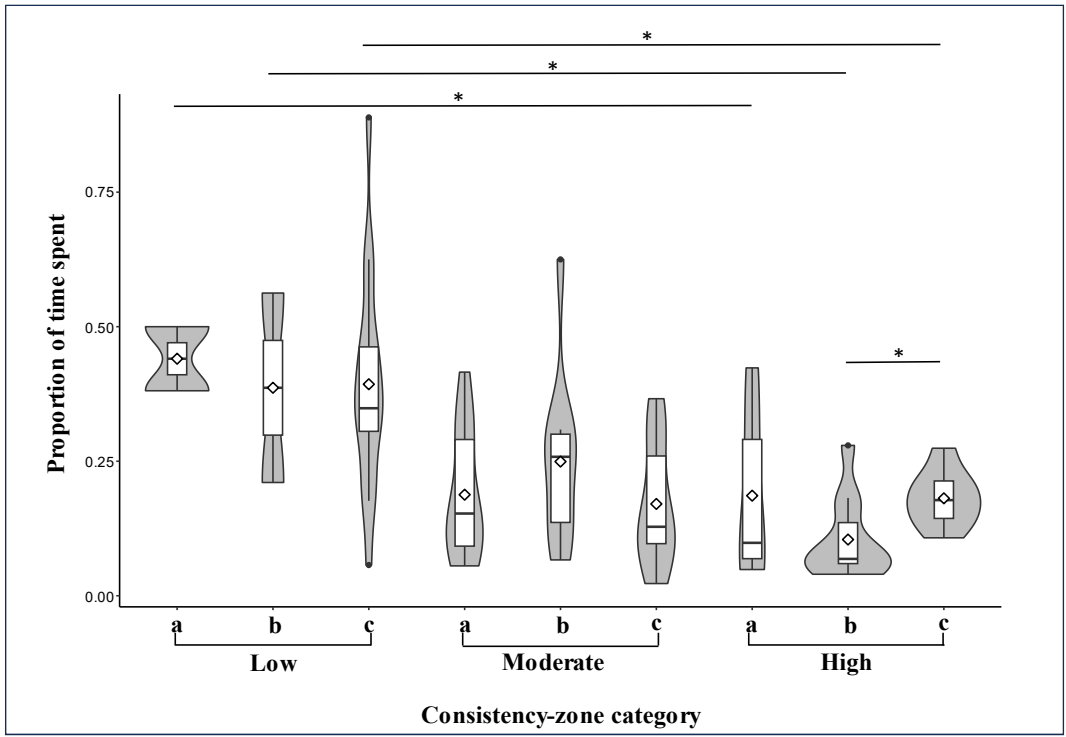

**Fig S7: Begging behavior :** Box plot embedded within violin plot describing the comparison among dogs from different consistency zones for begging behavior.

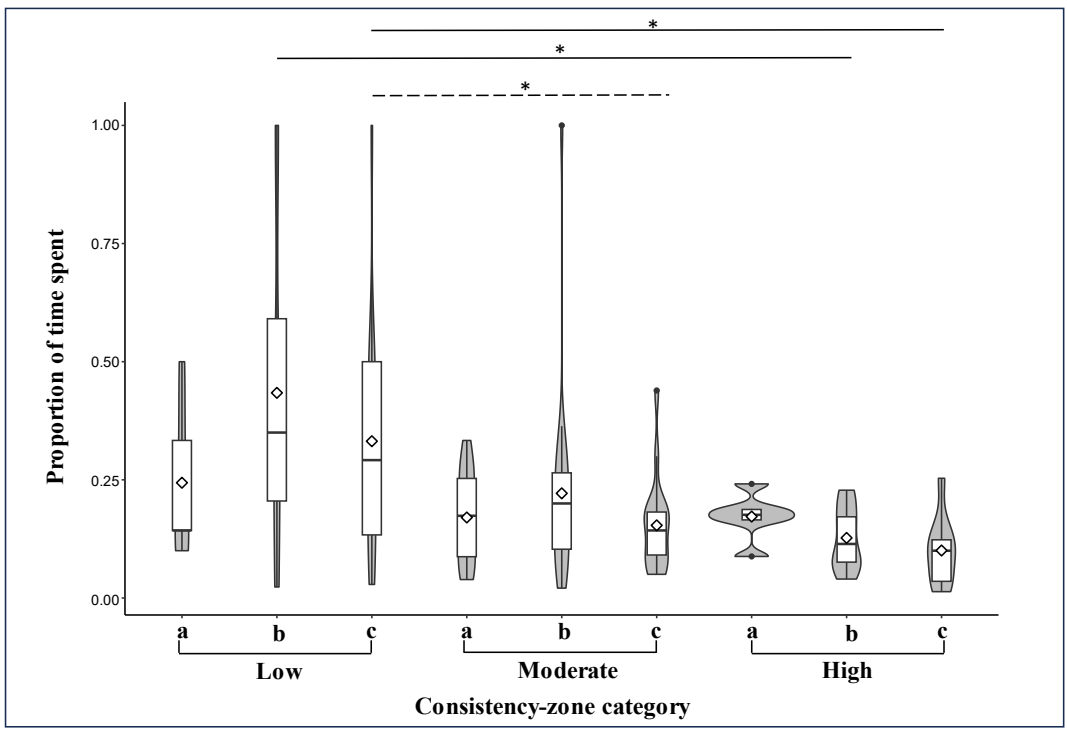

**Fig S8: Vigilance behavior :** Box plot embedded within violin plot describing the comparison among dogs from different consistency zones for vigilance behavior.

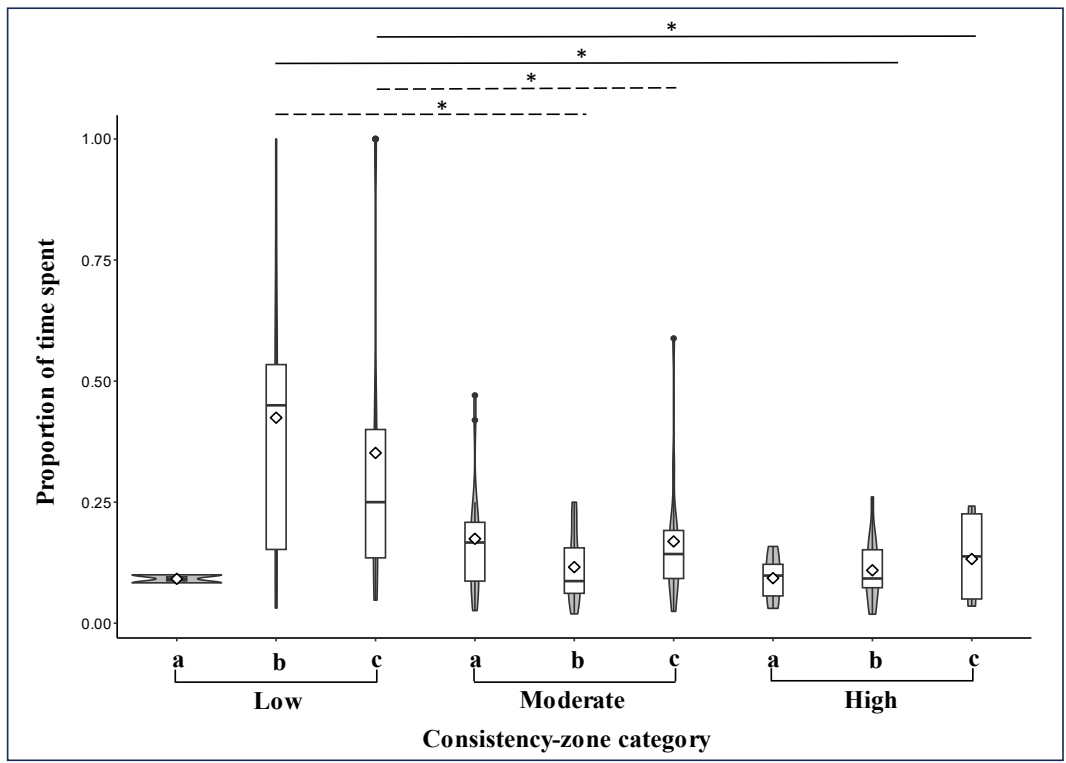

**Fig S9: Gait behavior:** Box plot embedded within violin plot describing the comparison among dogs from different consistency zones for gait behavior.